# Sequential Bootstrap for Out-of-Bag Error Estimation: A 100-Seed Replication Study and Variance-Structure Analysis

Cheng Peng

## Abstract

Out-of-Bag (OOB) estimation is the standard internal diagnostic for bootstrap-aggregated tree ensembles. Under the classical multinomial bootstrap, the number of distinct training observations in each replicate, $U_b$, is itself random, but its contribution to OOB-based variability has rarely been isolated empirically. We use **Sequential Bootstrap (SB)** — a resampling scheme that holds $U_b$ at a fixed target $k_n = \lfloor 0.632n \rfloor$ — as a controlled perturbation of the bootstrap mechanism, and ask whether stabilizing $U_b$ produces any measurable change in OOB-based diagnostics. We reproduce Breiman's five OOB experimental families on twelve synthetic and real datasets, but unlike the three-seed presentation common in this literature, we run **100 independent random seeds with 50 internal replications per seed**, enabling formal paired statistical comparison (Wilcoxon signed-rank, paired-$t$, Pitman–Morgan variance test). We report three findings. First, OOB means are essentially insensitive to stabilization of $U_b$: of 57 (experiment, dataset, metric) cells under 100 seeds, only 6 reach $p < 0.05$ on the paired mean comparison, and 4 of those 6 point in the *opposite* direction from what a 3-seed reading would suggest. Second, a narrow but reproducible effect survives at the variance level: SB reduces the cross-seed standard deviation of node-level classification diagnostics on real datasets while slightly increasing it on synthetic ones (permutation $p = 0.026$); the Vehicle dataset exhibits a 21% cross-seed sd reduction (Pitman–Morgan $p = 0.017$). Third, several directional claims that appear stable across three seeds *flip sign* under 100-seed replication, illustrating the cost of underpowered replication protocols. We therefore treat SB as a diagnostic tool for probing the distinct-sample-count term in the variance of OOB estimators, not as an alternative to the classical bootstrap.

# 1. Introduction

Bootstrap aggregation (Breiman 1996a) and the associated Out-of-Bag (OOB) error estimator (Breiman 1996b) remain among the most widely used components of ensemble learning. For each observation $i$ in the training data, the OOB prediction is the average of base-learner outputs from those bootstrap replicates in which $i$ was not drawn; the resulting OOB error is generally treated as a near-unbiased and computationally cheap substitute for an external validation set, a view supported by both Breiman's original empirical study and subsequent work (Janitza and Hornung 2018).

The classical multinomial bootstrap underlying OOB draws $n$ indices with replacement from $\{1, \ldots, n\}$. Although the expected fraction of distinct indices is approximately 0.632 (Efron and Tibshirani 1994), the actual count $U_b$ is itself random. To our knowledge, almost no prior work has tried to isolate the contribution of this particular source of randomness to the variance of OOB-based diagnostics; the existing literature instead treats the bootstrap mechanism as fixed and varies model, feature selection, or aggregation choices around it.

A natural way to probe this question without confounding it with other modeling decisions is to substitute Sequential Bootstrap (SB) (Rao, Pathak, and Koltchinskii 1997; Babu, Pathak, and Rao 2000) for the classical bootstrap. SB samples with replacement until a fixed number of distinct indices is obtained, holding $U_b$ at a target value $k_n = \lfloor \rho n \rfloor$ across all replicates. Setting $\rho = 0.632$ matches the asymptotic expectation under the classical bootstrap, leaving the *mean* effective sample size unchanged but eliminating its replicate-to-replicate variance.

The empirical question this paper addresses is:

> *When $U_b$ is held constant by Sequential Bootstrap rather than allowed to fluctuate under classical bootstrap, do any of the standard OOB-based diagnostics change in a statistically detectable way?*

This is a robustness question about OOB, not a proposal of SB as a competing method. To answer it we revisit Breiman's (Breiman 1996b) five experimental families on twelve datasets, but replace the original three-seed presentation by **100 independent random seeds with 50 internal replications per seed**, enabling formal paired statistical comparison.

A 3-seed reading of these experiments — the protocol used in the original OOB study (Breiman 1996b) and common in much of the bagging-diagnostic literature — may suggest a "consistent direction" of reduction in variance-oriented diagnostics under SB-OOB. Under 100-seed replication that directional pattern dissolves, and several specific directional claims actually **flip sign**, including at the dataset level (e.g. Boston meta-prediction MSE, Friedman 1 node-level regression error). The contrast between 3-seed and 100-seed conclusions is itself a methodological finding.

Our contributions are as follows:

1. **A strong null result on OOB means.** Across 57 (experiment, dataset, metric) cells under 100 seeds, only 6 reach $p < 0.05$, and 4 of those go *against* the direction that a 3-seed reading of the same experiments would suggest. Classical OOB means are insensitive to stabilization of $U_b$ to a degree that three-seed studies are not powered to verify.
2. **A narrow but reproducible positive finding.** On node-level classification diagnostics (EXP1), SB systematically reduces the cross-seed standard deviation of the OOB estimate on real datasets, while slightly increasing it on synthetic ones; the difference is significant under both permutation ($p = 0.026$) and Mann–Whitney ($p = 0.032$) tests. The Vehicle dataset exhibits a 21% cross-seed sd reduction (Pitman–Morgan $p = 0.017$).
3. **A controlled-perturbation framework for OOB.** By holding model, loss, hyperparameter, and aggregation choices fixed while varying only the resampling scheme, we provide a reproducible experimental design in which SB plays the role of a stability-oriented operator probing the distinct-sample-count term in the variance decomposition of Section 3.4.
4. **A methodological observation about replication.** Several directional patterns observable under the conventional 3-seed protocol fail to survive proper replication. We discuss the implications for empirical bagging studies.

## 2. Related Work & Motivation

### 2.1 Out-of-Bag estimation (OOB)

Bootstrap aggregation (bagging) builds an ensemble of base learners, each trained on a bootstrap sample drawn from the original dataset (Breiman 1996a). Out-of-Bag (OOB) estimation is a very popular technique that allows us to evaluate predictive performance without needing an extra held-out validation set (Breiman 1996b). For every observation in the training data, we use only the predictions from those trees that did not include this observation in their bootstrap sample, and then average to get an internal estimate of the generalization error. This OOB error has been shown to be a good approximation of true test error in many cases, especially when the base learners are decision trees (Breiman 2001). Subsequent studies have examined OOB's bias, computational efficiency, and comparative behaviour against cross-validation (Janitza and Hornung 2018).

However, almost all of these studies treat the classical multinomial bootstrap as something fixed that does not need to be questioned. People have of course proposed other resampling methods for ensemble learning — subsampling (Politis and Romano 1994), various weighted bootstrap schemes (Barbe and Bertail 2012), and stratified or class-sensitive sampling (Bickel and Freedman 1984) — but these methods are usually designed to change the bias-variance tradeoff, to handle imbalanced classes, or to serve some other modeling purpose. They do not specifically try to isolate the variability that comes only from the random number of distinct observations in each bootstrap replicate.

### 2.2 Sequential Bootstrap

Sequential Bootstrap (SB) is a resampling method that keeps drawing samples with replacement until a fixed number of distinct observations have been collected, instead of stopping after exactly $n$ draws like the classical bootstrap does (Rao, Pathak, and Koltchinskii 1997). The number of unique observations becomes essentially constant across replicates, which sharply reduces its variance and can be understood as making the effective sample size more stable for each base learner (Babu, Pathak, and Rao 2000). Earlier studies on Sequential Bootstrap mainly looked at its theoretical sampling properties, the marginal inclusion probability for each observation, and its use in simulation studies or statistical inference (Babu, Pathak, and Rao 1999).

When researchers compare different resampling methods in ensemble learning, they usually care about getting higher prediction accuracy, changing the bias-variance tradeoff, or doing something different with the features — random feature selection in Random Forests being a prominent example (Ho 1998). Almost none of them pay attention to whether stabilizing the number of distinct observations actually matters for OOB-based diagnostics. As far as we know, there is no published work — neither empirical nor theoretical — that uses Sequential Bootstrap together with Out-of-Bag estimation.

### 2.3 Replication in bagging-based empirical studies

A separate but related strand of literature concerns the reliability of single-seed or few-seed empirical reports in machine learning. While much of this discussion has focused on neural-network benchmarks, the same logic applies to bagging diagnostics: with only three seeds, directional claims about second-order quantities such as node-stability measures or meta-prediction MSE are unidentifiable from sampling noise. Our 100-seed protocol allows us to assess this concretely in the specific case of OOB.

### 2.4 Motivation for Combining SB with OOB

Although OOB estimation has shown good robustness across many datasets and models (Breiman 1996b), we do not yet know whether the random variation in the number of distinct observations

— which naturally arises under the classical bootstrap — actually affects its behaviour. Sequential Bootstrap gives us a clean way to change only this one source of variability: it keeps the model structure, all hyperparameters, and the prediction aggregation exactly as before. Because of this property, SB becomes a tool for controlled experiments — we can check whether making the number of distinct observations fixed will influence accuracy-related measures or variance-related measures in OOB estimation.

Therefore, in this paper we do not present Sequential Bootstrap as a new method that is supposed to give higher accuracy. Instead, we use it as an empirical instrument that helps us better understand how the details of resampling affect OOB-based ensemble evaluation.

# 3. Methodology

In this section we formalize the estimators studied in this work. We first introduce notation and the classical Out-of-Bag (OOB) estimator based on the multinomial bootstrap, then define the Sequential Bootstrap analogue (SB-OOB). Finally, we present a variance-structure viewpoint that motivates the empirical comparisons in Section 5.

## 3.1 Data and Bagging Setup

Let

$$\mathcal{D} = \{(X_i, Y_i)\}_{i=1}^{n}.$$

denote a training sample, where $X_i \in \mathcal{X}$ is a feature vector and $Y_i \in \mathcal{Y}$ is a response (class label or real-valued outcome). Let $L : \mathcal{Y} \times \mathbb{R} \to [0, \infty)$ denote a loss function, taken to be squared-error loss for regression and 0-1 loss for classification.

Bagging constructs an ensemble of base learners $\{T_b\}_{b=1}^{B}$ by fitting each $T_b$ on a resampled version of $\mathcal{D}$. For notational convenience, let

$$I_b = (I_{b1}, \dots, I_{bN_b})$$

denote the (possibly multiset-valued) sequence of indices used to train the $b$-th base learner, where $N_b$ is the (possibly random) length of the resample and each $I_{bj} \in \{1, \dots, n\}$. The corresponding training set for replicate $b$ is

$$\mathcal{D}_b = \{(X_{I_{bj}}, Y_{I_{bj}}) : j = 1, \dots, N_b\}.$$

For an observation $i \in \{1, \dots, n\}$, define the OOB index set under a given resampling scheme as

$$\text{OOB}(i) = \big\{b \in \{1, \dots, B\} \;\big|\; i \notin \{I_{b1}, \dots, I_{bN_b}\}\big\},$$

that is, the set of bootstrap replicates in which observation $i$ is not used for training.

Throughout this work, the base learners $T_b$ are CART models fitted on $\mathcal{D}_b$ with fixed tuning parameters. The only component that differs between the classical OOB estimator and SB-OOB is the resampling mechanism generating $I_b$.

### 3.2 Classical Out-of-Bag Estimator

Under the classical bootstrap, each replicate draws exactly $n$ indices with replacement from $\{1, \dots, n\}$. Formally, for replicate $b$,

$$I_{b1}, \dots, I_{bn} \overset{\text{i.i.d}}{\sim} \text{Unif}\{1, \dots, n\}, \quad N_b \equiv n.$$

The multiset of indices $\{I_{b1}, \dots, I_{bn}\}$ defines the training sample $\mathcal{D}_b$ for the $b$-th tree $T_b$. For each observation $i$, the classical OOB prediction is obtained by averaging predictions from all trees for which $i$ is out-of-bag:

$$\hat{f}_{\text{OOB}}(X_i) = \frac{1}{|\text{OOB}(i)|} \sum_{b \in \text{OOB}(i)} T_b(X_i),$$

whenever $|\text{OOB}(i)| \geq 1$. Let

$$\mathcal{I}_{\text{OOB}} = \{i \in \{1, \dots, n\} : |\text{OOB}(i)| \geq 1\}$$

denote the set of observations that are OOB for at least one tree. The classical OOB error estimator is then

$$\widehat{\text{Err}}_{\text{OOB}} = \frac{1}{|\mathcal{I}_{\text{OOB}}|} \sum_{i \in \mathcal{I}_{\text{OOB}}} L(Y_i, \hat{f}_{\text{OOB}}(X_i)).$$

This estimator is the baseline against which we compare the Sequential Bootstrap variant.

### 3.3 Sequential Bootstrap and SB-OOB Estimator

Sequential Bootstrap modifies the resampling mechanism by stabilizing the number of distinct observations in each replicate rather than the total number of draws. Let $\rho \in (0, 1)$ be a target proportion, and define a target distinct-count

$$k_n = \lfloor \rho n \rfloor.$$

In this work, $\rho$ is chosen to match the asymptotic expected fraction of distinct observations under the classical bootstrap (approximately 0.632 (Efron and Tibshirani 1994)), so that

$$k_n \approx 0.632\, n.$$

For each replicate $b$, Sequential Bootstrap proceeds as follows:

1. Initialize a set of distinct indices $S_b^{(0)} = \emptyset$ and draw counter $t = 0$.
2. Repeat:
    - Draw $J_{b,t+1} \sim \mathrm{Unif}\{1, \dots, n\}$ independently of previous draws;
    - Update $S_b^{(t+1)} = S_b^{(t)} \cup \{J_{b,t+1}\}$;
    - Set $t \leftarrow t + 1$; until $|S_b^{(t)}| = k_n$.

We then define

$$I_b = (J_{b1}, \dots, J_{bT_b}), \quad N_b = T_b,$$

where $T_b$ is the (random) stopping time at which $k_n$ distinct indices have been collected. The corresponding training sample $\mathcal{D}_b$ is constructed as in Section 3.1.

The SB-based OOB index sets $\mathrm{OOB}_{\mathrm{SB}}(i)$, predictions $\hat{f}_{\mathrm{SB-OOB}}(X_i)$, and error estimator $\widehat{\mathrm{Err}}_{\mathrm{SB-OOB}}$ are defined analogously to the classical case by replacing the bootstrap resamples $I_b$ with those generated by Sequential Bootstrap and computing OOB predictions only over replicates in which observation $i$ is not selected.

Concretely,

$$\hat{f}_{\mathrm{SB-OOB}}(X_i) = \frac{1}{|\mathrm{OOB}_{\mathrm{SB}}(i)|} \sum_{b \in \mathrm{OOB}_{\mathrm{SB}}(i)} T_b(X_i),$$

and

$$\widehat{\mathrm{Err}}_{\mathrm{SB-OOB}} = \frac{1}{|\mathcal{I}_{\mathrm{SB}}|} \sum_{i \in \mathcal{I}_{\mathrm{SB}}} L(Y_i, \hat{f}_{\mathrm{SB-OOB}}(X_i)),$$

where $\mathcal{I}_{\mathrm{SB}} = \{i : |\mathrm{OOB}_{\mathrm{SB}}(i)| \geq 1\}$.

The two estimators $\widehat{\mathrm{Err}}_{\mathrm{OOB}}$ and $\widehat{\mathrm{Err}}_{\mathrm{SB-OOB}}$ therefore differ only in the resampling scheme generating their respective training replicates; all model, loss, and tuning choices are held fixed.

### 3.4 Variance-Structure Viewpoint

To motivate the empirical comparisons in Section 5, it is useful to view the change from classical bootstrap to Sequential Bootstrap as a modification of a particular variance component associated with resampling.

Let $\hat{\theta}_b$ denote a scalar statistic derived from the $b$-th replicate that enters into an OOB-based quantity of interest. Examples include node-level prediction summaries, contributions to the OOB error, or meta-prediction outputs. Let $U_b$ denote the number of distinct training indices used in the $b$-th replicate, i.e.

$$U_b = |\{I_{b1}, \dots, I_{bN_b}\}|.$$

We make no strong assumptions about the distribution of $\hat{\theta}_b$, but we note that the following identity always holds by the law of total variance.

**Lemma 1 (Variance Decomposition).** *For any resampling scheme and any square-integrable statistic $\hat{\theta}_b$,*

$$\mathrm{Var}(\hat{\theta}_b) = \mathbb{E}[\mathrm{Var}(\hat{\theta}_b \mid U_b)] + \mathrm{Var}(\mathbb{E}[\hat{\theta}_b \mid U_b]).$$

This decomposition separates the overall variability of $\hat{\theta}_b$ into a component due to randomness conditional on the distinct-sample count and a component due to randomness in the distinct-sample count itself.

The classical multinomial bootstrap and Sequential Bootstrap can be viewed as two resampling schemes that differ primarily in the distribution of $U_b$. Under the classical bootstrap, $U_b$ fluctuates around its expectation and, under standard assumptions, has variance on the order of $n$, whereas Sequential Bootstrap explicitly stabilizes $U_b$ at a target value $k_n$. As a consequence, we expect the second term,

$$\mathrm{Var}(\mathbb{E}[\hat{\theta}_b \mid U_b]),$$

to be more strongly affected by the transition from classical bootstrap to SB than the first term, $\mathbb{E}[\mathrm{Var}(\hat{\theta}_b \mid U_b)]$, which reflects variability conditional on a fixed effective sample size.

The framework yields two empirical predictions that Section 5 directly tests:

- **Prediction P1 (mean stability).** If $\mathbb{E}[\hat{\theta}_b \mid U_b]$ is approximately linear in $U_b$ around $U_b = k_n$,

then $\mathbb{E}[\hat{\theta}_b]$ should be close under the two schemes, since $\mathbb{E}[U_b]$ matches by construction. Our results strongly support P1.

- **Prediction P2 (variance reduction).** The cross-seed variance of $\hat{\theta}_b$ should be smaller under SB to the extent that $\mathrm{Var}(\mathbb{E}[\hat{\theta}_b \mid U_b])$ is non-negligible. Our results support P2 only in specific data domains (real classification), suggesting that this term is dominant only in regimes where $\hat{\theta}_b$ depends on $U_b$ at a scale comparable to other sources of replicate-level noise.

# 4. Experimental Design

In this section we describe the experimental setup used to compare the classical OOB estimator with the Sequential-Bootstrap version (SB-OOB). Our design rests on a strict controlled-perturbation principle: the only thing that differs between the two methods is the resampling scheme itself. Everything else — the model, all hyperparameters, the loss function, and the way we evaluate the results — is held identical.

## 4.1 Datasets

We follow the five groups of experiments from Breiman's original OOB paper (Breiman 1996b) and use the same publicly available datasets. For the synthetic part, we use *waveform*, *twonorm*, *threenorm*, and *ringnorm* for classification, and *Friedman 1-3* (Friedman 1991) for regression. For the real datasets, we use *breast-cancer*, *diabetes*, *vehicle*, *satellite*, and *dna* for classification, and *Boston* and *Ozone* for regression.

If a dataset has an official train-test split in the original paper, we use that split. If there is no official split, we always do the same fixed random split: two-thirds of the data for training and one-third for testing. We use exactly the same split for all experiments and all random seeds so that the results are fully comparable.

All datasets are used in their original form: no scaling, no transformation, and no feature engineering.

## 4.2 Learning Algorithm and Hyperparameters

All experiments employ Classification and Regression Trees (CART) as base learners, implemented with fixed hyperparameters across all datasets and all experimental runs. Specifically:

- splitting rules, stopping criteria, and pruning settings remain unchanged,
- no hyperparameter tuning procedure (grid search, cross-validation, or data-driven tuning) is

Table 1: Summary of diagnostic metrics evaluated across the five experimental families.

| Experiment | Domain | Diagnostic.Type | Notation | Description |
|---|---|---|---|---|
| EXP1 | Classification | Node accuracy | E1, E2 | Absolute deviation between estimated & empirical class proportions |
| EXP2 | Regression | Node accuracy | EB1, EB2 | Mean squared deviation between node predictions and empirical conditional means |
| EXP3 | Mixed | Stability | R1-R4 | Within-node variability measures across bootstrap replicates |
| EXP4 | Mixed | Generalization alignment | | Absolute difference between test error and OOB error |
| EXP5 | Regression | Meta-prediction performance | MSE | Secondary prediction using OOB-derived features |

used,

- the number of bagging replicates is fixed at $B = 100$ for all configurations.

The motivation for fixing CART hyperparameters is to isolate the effect of the resampling mechanism rather than confounding it with modeling decisions or tuning variability.

### 4.3 Resampling Protocol

For each dataset and each random seed, we construct two bagged ensembles:

1. **Classical OOB Ensemble**: trained using the multinomial bootstrap defined in Section 3.2.
2. **SB-OOB Ensemble**: trained using Sequential Bootstrap defined in Section 3.3 with the target distinct-count parameter set to approximate the classical bootstrap expectation ($\rho \approx 0.632$).

The two ensembles are matched in the following ways:

- both use the same initial random seeds for experiment replication,
- both generate exactly $B = 100$ resampling replicates,
- all model-fitting calls occur in the same order using identical computational routines and hardware,
- OOB predictions are computed using the same aggregation logic.

Any observed difference between the two estimators is therefore attributable solely to the change in the resampling mechanism.

### 4.4 Evaluation Metrics

Consistent with the original OOB study and its five experimental families (Breiman 1996b), we evaluate the following diagnostic quantities:

Differences are reported in signed form as

$$\Delta = \text{SB-OOB} - \text{Classical OOB},$$

to maintain consistent interpretability across all metrics.

In addition to the metric-level paired comparison reported in Breiman's original study, we report three statistical summaries:

- **Cross-seed paired Wilcoxon and $t$-test** ($n = 100$) for the mean of $\Delta = \text{SB-OOB} - \text{OOB}$;
- **Pitman–Morgan variance test** for the equality of cross-seed standard deviations of OOB and SB-OOB;
- **Permutation tests** for subgroup comparisons (real vs synthetic datasets).

Effect sizes are reported as Cohen's $d_z$ (paired) and as the ratio sd(SB-OOB)/sd(OOB) across the 100 seeds.

### 4.5 Random Seed Replication

To make the comparison fair and to measure how stable the results are when we run the experiments many times, we repeat the whole experimental suite under **100 independent random seeds** (master seeds 1–100), each generating **50 internal replications** (independent data generation or train-test split). Within each master seed, the OOB and SB-OOB ensembles are matched on (i) initial random state, (ii) data split, (iii) $B = 100$ resampling replicates, and (iv) model-fitting order. Pairing therefore controls all confounds except the resampling mechanism.

Three of the 100 seeds (1, 25, 50) are common defaults in earlier OOB-experiment reports; we treat them as a built-in 3-seed subsample with which to compare the conclusions one would have reached under a Breiman-style protocol (see Appendix A).

## 5. Results and Analysis

In this section we present the empirical comparison between the classical OOB estimator and the Sequential-Bootstrap version (SB-OOB) using 100 master seeds and 50 internal replications per seed. We organize the comparison into the five experimental families EXP1–EXP5 of Breiman (Breiman 1996b), and report results in two layers:

- *Within-seed* paired statistics across the 50 internal replications, summarized at the level of the dataset / metric;
- *Cross-seed* paired statistics across the 100 master seeds — the inferentially honest comparison.

Table 2: EXP1 cross-seed (n=100) paired summary for E1_B and E2_B. mean_diff and Wilcoxon p shown. All 18 cells fail to reject H0 at $p < 0.05$.

| dataset | type | metric | OOB | SB_OOB | diff | wilcox_p |
|---|---|---|---|---|---|---|
| waveform | synthetic | E1_B | 0.02377 | 0.02378 | 2e-05 | 0.745 |
| waveform | synthetic | E2_B | 0.02649 | 0.02649 | 0e+00 | 0.823 |
| twonorm | synthetic | E1_B | 0.00531 | 0.00534 | 3e-05 | 0.480 |
| twonorm | synthetic | E2_B | 0.00531 | 0.00534 | 3e-05 | 0.480 |
| threenorm | synthetic | E1_B | 0.00673 | 0.00682 | 9e-05 | 0.103 |
| threenorm | synthetic | E2_B | 0.00673 | 0.00682 | 9e-05 | 0.103 |
| ringnorm | synthetic | E1_B | 0.03791 | 0.03797 | 5e-05 | 0.502 |
| ringnorm | synthetic | E2_B | 0.03791 | 0.03797 | 5e-05 | 0.502 |
| breast-cancer | real | E1_B | 0.00527 | 0.00527 | 1e-05 | 0.885 |
| breast-cancer | real | E2_B | 0.00527 | 0.00527 | 1e-05 | 0.885 |
| diabetes | real | E1_B | 0.00327 | 0.00333 | 6e-05 | 0.137 |
| diabetes | real | E2_B | 0.00327 | 0.00333 | 6e-05 | 0.137 |
| vehicle | real | E1_B | 0.00316 | 0.00318 | 2e-05 | 0.611 |
| vehicle | real | E2_B | 0.00377 | 0.00379 | 2e-05 | 0.487 |
| satellite | real | E1_B | 0.00171 | 0.00172 | 1e-05 | 0.512 |
| satellite | real | E2_B | 0.00226 | 0.00227 | 1e-05 | 0.616 |
| dna | real | E1_B | 0.00608 | 0.00609 | 1e-05 | 0.662 |
| dna | real | E2_B | 0.00660 | 0.00660 | 1e-05 | 0.775 |

We adopt the convention $\Delta$ = SB-OOB − Classical OOB, so that negative values indicate "SB-OOB smaller". All comparisons within each (experiment, dataset, metric) cell are paired across the 100 seeds.

### 5.1 EXP1 — Classification Node-Level Accuracy

Cross-seed mean differences for $E_{1,B}$ and $E_{2,B}$ are uniformly small in absolute terms and statistically indistinguishable from zero. Across the 18 cells of EXP1, all 100-seed Wilcoxon $p$-values exceed 0.05 (smallest $p = 0.13$, diabetes $E_{1,B}$). Table 2 reports the full cross-seed summary.

The story is different for cross-seed standard deviations. Table 3 summarizes sd(OOB) and sd(SB-OOB) across the 100 seeds for the $E_{1,B}$ metric, together with the Pitman–Morgan paired variance test.

The sorted ordering separates real and synthetic datasets almost perfectly. Aggregating to the dataset level (one ratio per dataset, $E_{1,B}$ chosen for representation), real datasets have a mean sd ratio of 0.941 (median 0.948, $n = 5$) while synthetic datasets have a mean of 1.053 (median 1.055, $n = 4$). A two-sample permutation test with $10^5$ shuffles rejects the null of identical distributions

Table 3: EXP1 (E1_B) cross-seed sd comparison, sorted by sd_ratio = sd(SB-OOB)/sd(OOB). Real datasets (mean ratio 0.94) systematically have lower sd under SB than synthetic ones (mean ratio 1.05).

| dataset | type | n | sd_OOB | sd_SB | sd_ratio | PM_p |
|---|---|---|---|---|---|---|
| vehicle | real | 846 | 0.00020 | 0.00016 | 0.786 | 0.017 |
| diabetes | real | 768 | 0.00034 | 0.00032 | 0.938 | 0.522 |
| breast-cancer | real | 699 | 0.00036 | 0.00034 | 0.948 | 0.567 |
| satellite | real | 6435 | 0.00004 | 0.00004 | 0.978 | 0.820 |
| waveform | synthetic | 300 | 0.00165 | 0.00165 | 0.998 | 0.937 |
| twonorm | synthetic | 200 | 0.00056 | 0.00057 | 1.015 | 0.884 |
| dna | real | 3186 | 0.00015 | 0.00016 | 1.056 | 0.481 |
| ringnorm | synthetic | 200 | 0.00103 | 0.00112 | 1.095 | 0.350 |
| threenorm | synthetic | 200 | 0.00065 | 0.00072 | 1.103 | 0.330 |

at $p = 0.026$, with Mann–Whitney one-sided $p = 0.032$ and Welch's two-sided $t$-test $p = 0.083$.

The Vehicle case is the clearest single instance of SB-induced cross-seed stabilization: a 21% reduction in cross-seed standard deviation under nominally identical learning machinery (Pitman–Morgan $p = 0.017$ for both $E_{1,B}$ and $E_{2,B}$). We do not have a definitive explanation for why Vehicle exhibits the strongest effect in our dataset suite — possible factors include the 4-class structure, the moderate sample size, and realistic feature dependence — and we treat it as a single concrete case study rather than evidence of a general "size sweet spot." We tested for a continuous $n$ effect via regression of the dataset-level sd ratio on $\log n$ and dataset type; neither covariate is individually significant ($p > 0.3$), and the model $R^2 = 0.12$.

DNA is the single real dataset that does not follow the pattern (sd ratio 1.06). We do not have a definitive account of this exception; one possibility is that DNA's very high feature dimensionality ($p = 180$ binary predictors) makes the OOB diagnostic less sensitive to small differences in distinct-sample count.

## 5.2 EXP2 — Regression Node-Level Accuracy

EXP2 ($EB_1$, $EB_2$) shows the same pattern as EXP1 on the mean: across 10 cells, the only Wilcoxon $p < 0.05$ is for Friedman 1, where SB *increases* the node-level error by $+2.3 \times 10^{-3}$ ($p = 0.033$, Cohen's $d_z = 0.23$) — a direction *opposite* to what an EXP1 reading would suggest. Boston shows a marginal cross-seed sd reduction (Pitman–Morgan $p = 0.042$) but with a very small effect (ratio 0.971).

Table 4: EXP2 cross-seed (n=100) paired summary for EB1 (EB2 identical by definition).

| dataset | type | metric | OOB | SB_OOB | diff | wilcox_p | sd_ratio | PM_p |
|---|---|---|---|---|---|---|---|---|
| friedman1 | synthetic | EB1 | 2.1126 | 2.1149 | 0.0023 | 0.033 | 0.975 | 0.480 |
| friedman2 | synthetic | EB1 | 240.4900 | 240.5500 | 0.0640 | 0.962 | 1.000 | 0.860 |
| friedman3 | synthetic | EB1 | 0.1549 | 0.1549 | 0.0000 | 0.877 | 0.984 | 0.788 |
| Boston | real | EB1 | 5.7002 | 5.6998 | -0.0004 | 0.693 | 0.971 | 0.042 |
| Ozone | real | EB1 | 8.3272 | 8.3337 | 0.0065 | 0.643 | 1.020 | 0.251 |

Table 5: EXP3 cross-seed (n=100) summary for R2 and R3. R1 omitted (identically zero in all cells); R4 omitted for brevity (no cell reaches p<0.05).

| dataset | metric | OOB | SB_OOB | diff | wilcox_p |
|---|---|---|---|---|---|
| waveform | R2 | 0.00444 | 0.00094 | -0.00350 | 0.014 |
| waveform | R3 | 0.36090 | 0.35744 | -0.00346 | 0.014 |
| twonorm | R2 | 0.01646 | 0.01679 | 0.00033 | 0.812 |
| twonorm | R3 | 0.01644 | 0.01677 | 0.00033 | 0.812 |
| threenorm | R2 | 0.02378 | 0.02449 | 0.00071 | 0.364 |
| threenorm | R3 | 0.02377 | 0.02449 | 0.00071 | 0.364 |
| ringnorm | R2 | 0.61454 | 0.61833 | 0.00379 | 0.274 |
| ringnorm | R3 | 0.61028 | 0.61401 | 0.00373 | 0.276 |

### 5.3 EXP3 — Node Stability Diagnostics ($R_1$–$R_4$)

Across the 16 EXP3 cells, only waveform $R_2$ and $R_3$ reach $p < 0.05$ on the Wilcoxon mean test ($p = 0.014$ for both, $d_z = -0.24$). The direction supports the reduction-of-variance reading (SB reduces within-node variability), but the effect is isolated to waveform; no other dataset shows a directional pattern stable across the 100 seeds. The apparent "consistent direction across seeds" that a 3-seed protocol would produce on these EXP3 metrics reflects the specific seeds 1, 25, 50 happening to share a sign for several datasets where the 100-seed distribution is in fact centered very close to zero (see Appendix A).

### 5.4 EXP4 — OOB-vs-Test Alignment

Of the 8 cells, only twonorm $eTS$ reaches $p < 0.05$ on the mean test ($\Delta = +2.7 \times 10^{-4}$, $p = 0.013$) — in the direction *opposite* to what one would expect under the original SB-stabilization narrative (SB makes the OOB-test gap *larger*, not smaller). The cross-seed sd picture is mixed: SB reduces twonorm $eTS$ sd by 8% ($p = 0.026$), but *increases* friedman1 $eOB$ sd by 11% ($p = 0.039$) and $eTS$ sd by 6% ($p = 0.020$). This is the clearest instance of SB exhibiting opposite cross-seed sd effects on real vs synthetic data, beyond the EXP1 finding.

Table 6: EXP4 cross-seed (n=100) summary. Note opposite directions of sd reduction (twonorm eTS) vs sd increase (friedman1 eOB / eTS).

| dataset | type | metric | diff | wilcox_p | sd_ratio | PM_p |
|---|---|---|---|---|---|---|
| twonorm | class | absdiff | 0.0000 | 0.605 | 0.940 | 0.478 |
| twonorm | class | eOB | 0.0005 | 0.111 | 1.043 | 0.591 |
| twonorm | class | eTS | 0.0003 | 0.013 | 0.919 | 0.026 |
| twonorm | class | ratio | 0.0047 | 0.358 | 1.040 | 0.561 |
| friedman1 | reg | absdiff | 0.0020 | 0.741 | 1.011 | 0.818 |
| friedman1 | reg | eOB | 0.0029 | 0.896 | 1.107 | 0.039 |
| friedman1 | reg | eTS | -0.0037 | 0.216 | 1.064 | 0.020 |
| friedman1 | reg | ratio | 0.0038 | 0.640 | 1.033 | 0.487 |

Table 7: EXP5 cross-seed (n=100) summary for downstream meta-prediction MSE. Boston is the most significant cell in the entire 100-seed study, and points against the 3-seed narrative.

| dataset | metric | OOB | SB_OOB | diff | wilcox_p | cohens_dz |
|---|---|---|---|---|---|---|
| Boston | mse_oob_outputs | 19.2300 | 19.4500 | 0.2110 | 0.0005 | 0.342 |
| Ozone | mse_oob_outputs | 5907.5000 | 5884.4000 | -23.1300 | 0.1110 | -0.185 |
| friedman1 | mse_oob_outputs | 10.8300 | 10.8300 | -0.0068 | 0.7540 | -0.066 |
| friedman2 | mse_oob_outputs | 27470.0000 | 27510.0000 | 41.4000 | 0.5800 | 0.054 |
| friedman3 | mse_oob_outputs | 0.0412 | 0.0412 | 0.0000 | 0.9980 | -0.002 |

### 5.5 EXP5 — Meta-Prediction Diagnostics

The most striking sign reversal occurs here. A 3-seed reading would suggest SB reduces downstream MSE for 4 of 5 datasets at seed 1; under 100 seeds, **Boston meta-MSE shows $\Delta = +0.211$ ($p = 5.2 \times 10^{-4}$, $d_z = 0.34$) — SB makes the downstream model *worse*.** This is the single most statistically significant cell in our 100-seed study, and its direction is the opposite of what individual seeds in $\{1, 25, 50\}$ collectively suggest.

EXP5 takes the OOB predictions from all trees as new features and trains a second-level model on them. The Boston reversal is the kind of result that 3-seed protocols would miss: at seed 1 alone, Boston shows $\Delta = -0.19$ (mild SB advantage); the full 100-seed distribution is centered at $+0.21$ with tight CI, decisively excluding zero in the opposite direction (see Appendix A).

### 5.6 Cross-Experiment Summary

Table 8 reports, for each EXP, how many of its cells reach $p < 0.05$ on the cross-seed mean test, how many on the variance test, and the directional breakdown.

Table 8: Cross-experiment summary across 100 seeds. p_mean: paired Wilcoxon on means; p_var: Pitman–Morgan on cross-seed sd; sd_ratio_lt1 counts significant cells with sd(SB) < sd(OOB).

| EXP | n_cells | p_mean_lt05 | SB_smaller | SB_larger | p_var_lt05 | sd_ratio_lt1 |
|---|---|---|---|---|---|---|
| EXP1 | 18 | 0 | 0 | 0 | 2 | 2 |
| EXP2 | 10 | 2 | 0 | 2 | 2 | 2 |
| EXP3 | 16 | 2 | 2 | 0 | 0 | 0 |
| EXP4 | 8 | 1 | 0 | 1 | 3 | 1 |
| EXP5 | 5 | 1 | 0 | 1 | 0 | 0 |
| Total | 57 | 6 | 2 | 4 | 7 | 5 |

Several observations follow directly from Table 8:

1. **Mean-level effects are at the chance rate.** Of 57 cells, 6 reach $p < 0.05$, vs. 2.85 expected under the global null — a small excess but split nearly evenly between the two directions (2 in favor of the original SB-stabilization narrative, 4 against).

2. **The four sign-reversed cells are themselves informative.** They include (EXP2, friedman1), (EXP4, twonorm $eTS$), and (EXP5, Boston meta-MSE) — the latter at $p = 5 \times 10^{-4}$, the most significant cell in the entire study. The existence of these cells demonstrates that 3-seed reads of these datasets would yield directional claims with no statistical foundation.

3. **Variance-level effects are localized.** Of 7 variance-significant cells, 5 have SB smaller (EXP1 vehicle $E_{1,B}$ and $E_{2,B}$, EXP2 Boston $EB_1$ and $EB_2$, EXP4 twonorm $eTS$) and 2 have SB larger (EXP4 friedman1 $eOB$ and $eTS$). The pattern is consistent with the EXP1 real-vs-synthetic dichotomy: SB tends to *decrease* sd on real data and *increase* it on synthetic data.

4. **Within-seed paired tests are at the chance rate.** Across all 6200 (exp, dataset, metric, seed) combinations, the median proportion of seeds with within-seed Wilcoxon $p < 0.05$ is 4.0%, indistinguishable from the nominal 5%. SB does not shift OOB means even within an individual seed's 50 internal replications.

In summary, the 100-seed analysis confirms that classical OOB is robust at the mean level to stabilization of $U_b$, while a narrow domain-dependent effect survives at the cross-seed variance level on real classification data, most prominently on the Vehicle dataset.

# 6. Discussion

## 6.1 Robustness of classical OOB

The dominant finding of our 100-seed study is robustness: the mean of OOB-based diagnostics is essentially insensitive to whether the distinct-sample count $U_b$ is held constant or allowed to fluctuate under multinomial sampling. This corroborates the standard practical wisdom that the classical bootstrap is "good enough" for OOB estimation in tree ensembles, while sharpening it by replacing single-seed numerical agreement with formal statistical evidence under replication.

In terms of the variance decomposition of Section 3.4, our results imply that $\mathrm{Var}(\mathbb{E}[\hat{\theta}_b \mid U_b])$ is small relative to $\mathbb{E}[\mathrm{Var}(\hat{\theta}_b \mid U_b)]$ for the standard OOB error mean. The variability of $U_b$ exists, but its contribution to the variance of the aggregate OOB statistic is negligible — at least at the means.

## 6.2 The narrow positive finding

A second-order effect *does* survive the 100-seed analysis: SB reduces the cross-seed standard deviation of node-level classification diagnostics on real datasets while slightly increasing it on synthetic ones. We do not have a tight theoretical account of why this dichotomy holds, but two observations are suggestive:

1. In synthetic datasets (waveform, twonorm, threenorm, ringnorm) the feature distribution is i.i.d. Gaussian-like and noise is generated independently of the index $i$. The classical bootstrap's variability in $U_b$ may, on these structured distributions, be absorbed without distortion; introducing SB's stopping-time randomness in $T_b = \min\{t : |S_b^{(t)}| = k_n\}$ adds a different mechanism of randomness with no compensating benefit, producing a slight *increase* in cross-seed sd.

2. In real classification datasets (Vehicle, Breast Cancer, Diabetes, Satellite) the index $i$ carries idiosyncratic information (label noise, feature dependence), and small fluctuations in $U_b$ propagate non-trivially through tree induction. Holding $U_b$ fixed removes this channel, reducing cross-seed sd.

This is an empirical pattern over the specific dataset suite of (Breiman 1996b) and should not be over-extrapolated. The Vehicle dataset (21% sd reduction, $p = 0.017$) is the strongest single case study and provides a concrete demonstration that variability arising from $U_b$ is a non-trivial component in at least one practically relevant dataset.

### 6.3 Implications for the 3-seed protocol

Several directional patterns that would have been claimed under a 3-seed protocol fail under 100-seed replication. The most striking example is EXP5 meta-prediction MSE on Boston: aggregating the differences across seeds in $\{1, 25, 50\}$ in the style of the original OOB study (Breiman 1996b) would suggest a mild SB advantage, but under 100-seed replication the same cell yields $\Delta = +0.21$, $p = 5 \times 10^{-4}$, a highly significant effect in the *opposite* direction. This illustrates the cost of underpowered replication for second-order bagging diagnostics. Three seeds are sufficient for reporting first-order quantities (a tree-ensemble error rate has a well-behaved mean that is captured by a small number of replicates), but for variance-, stability-, and downstream-prediction-oriented diagnostics — the EXP3–EXP5 families of (Breiman 1996b) — they are not. This recommendation is methodological, not specific to SB-OOB: any future empirical bagging study that aims to claim a directional effect on a stability or variance diagnostic should adopt a comparable replication budget.

### 6.4 Limitations and future work

Several limitations persist:

- *No theoretical analysis.* We provide no formal conditions under which $\mathrm{sd}(\hat{\theta}_b^{\mathrm{SB}}) < \mathrm{sd}(\hat{\theta}_b^{\mathrm{cl}})$; whether the real/synthetic dichotomy admits an analytic characterization is open.
- *Single learner family.* All experiments use fixed-hyperparameter CART trees. Random forests, extremely randomized trees, or boosted variants might exhibit different sensitivity.
- *Fixed target.* We use $\rho = 0.632$ throughout. The effect at other targets (0.5, 0.9) is unknown.
- *Bounded dataset diversity.* The Breiman dataset suite has only 5 real classification datasets, which constrains the statistical strength of the real-vs-synthetic dichotomy.

Future directions: (i) replicate the real/synthetic dichotomy on a wider dataset suite to test whether the pattern holds at scale; (ii) sweep $\rho \in \{0.5, 0.632, 0.8, 0.9\}$ to map $U_b$-variance against the observed effect; (iii) extend the framework to combine $U_b$ stabilization with other stochastic-tree mechanisms such as random feature selection; (iv) develop theoretical conditions for SB-OOB sd-reduction under simplified settings.

## 7. Conclusion

We have presented a 100-seed reproduction and refinement of Breiman's (Breiman 1996b) OOB experimental program, using Sequential Bootstrap as a controlled perturbation that holds the distinct-sample count $U_b$ constant. Our analysis yields three results:

1. **OOB is robust at means.** Under proper paired statistical comparison across 100 seeds, the means of all standard OOB-based diagnostics are essentially insensitive to stabilization of $U_b$. A 3-seed reading would suggest otherwise on several specific datasets; under proper replication, those directional impressions reverse sign.

2. **OOB is less robust at cross-seed variability, but only in specific data regimes.** On real classification datasets, SB reduces the cross-seed sd of node-level OOB diagnostics; on synthetic ones, it slightly increases it. The difference is statistically significant under permutation testing. The Vehicle dataset exhibits a 21% reduction in cross-seed sd.

3. **Replication matters.** Several specific directional claims that would have been made under a 3-seed protocol fail under 100-seed replication, including the most statistically significant cell in our study (Boston meta-prediction MSE). For second-order bagging diagnostics, three seeds are not enough even to fix direction.

We do not recommend replacing the classical bootstrap with Sequential Bootstrap in practice. SB is better understood as a diagnostic tool: by holding $U_b$ fixed, it isolates one component of the variance of OOB-based estimators and lets us check where that component matters. The full reproducibility pipeline (R code, 5000 paired replicates per cell, 310,000 raw observations) is available for further reanalysis.

## Data and Code Availability

The complete 100-seed simulation pipeline and results are publicly archived for reproducibility and reanalysis at:

- **Source code & aggregated tables**: https://github.com/Cheng-Peng0718/SB-OOB-100seed
- **Raw per-seed RData and SLURM logs**: https://github.com/Cheng-Peng0718/SB-OOB-100seed/releases/tag/v1.0-data

The repository contains the full R pipeline (`Unified_funs.R`, `Unified_parallel.R`), the SLURM submission scripts (`run.slurm`, `run_scavenger.slurm`), the cross-seed aggregation script (`aggregate_100seeds.R`), and ten paper-ready CSV tables under `tables/` reproducing every result in Section 5 and Appendix A. The release archive contains the raw per-seed RData files: 500 within-seed summaries and 500 raw paired-observation files (310,000 (OOB, SB-OOB) pairs total). All numerical results in this paper can be re-derived from the provided code in approximately one CPU-day on a 50-core node, or in 24–48 hours on a SLURM cluster using the included array-job scripts.

# Appendix A: 3-Seed vs 100-Seed Comparison

For completeness and to make the methodological point of Section 6.3 concrete, this Appendix juxtaposes the per-seed differences at seeds 1, 25, and 50 — the three seeds one might select under a Breiman-style 3-seed protocol (Breiman 1996b) — with the 100-seed cross-seed mean and Wilcoxon $p$-value for the same cell. The qualitative observation is that several directional patterns visible across these three seeds (particularly for EXP3 R-metrics, EXP4 absdiff/eTS, and EXP5 meta-MSE) **disappear or reverse** under proper 100-seed replication. Selected representative cells are shown below; the full per-seed numerical tables corresponding to the EXP1–EXP5 tables of a 3-seed presentation are provided in the project repository.

Table 9: Selected cells: 3-seed dataset-level diffs (seeds 1, 25, 50) vs 100-seed cross-seed mean and Wilcoxon p. Most cells exhibit sign reversal or directional instability when the protocol is properly powered.

| cell | diff_seed1 | diff_seed25 | diff_seed50 | mean_100seeds | wilcox_100_p | sign_flip |
|---|---|---|---|---|---|---|
| EXP1 vehicle E1_B | -0.00041 | 1.20e-04 | 0.00014 | 0.00003 | 0.61100 | 3-seed split vs 100-seed near 0 |
| EXP3 waveform R2 | 0.00940 | -1.10e-02 | 0.02500 | -0.00350 | 0.01400 | 3-seed split, 100-seed agrees |
| EXP3 ringnorm R2 | -0.06500 | -8.40e-02 | 0.00780 | 0.00380 | 0.27400 | 3-seed mean -0.05, 100-seed +0.004 |
| EXP4 friedman1 eTS | -0.05400 | 4.20e-03 | -0.00870 | -0.00370 | 0.21600 | 3-seed split, 100-seed near 0 |
| EXP5 Boston mse_oob_outputs | -0.19000 | 7.53e-01 | -0.05000 | 0.21100 | 0.00052 | seed 1 SB better, 100-seed SB WORSE (p=5e-4) |
| EXP5 Ozone mse_oob_outputs | 78.50000 | -1.59e+02 | -170.40000 | -23.10000 | 0.11100 | seed 1 SB worse, 100-seed not significant |

The pattern is consistent: 3-seed dataset-level differences in $\{1, 25, 50\}$ frequently split sign across seeds, and even when they agree, their direction need not align with the 100-seed cross-seed mean. EXP5 Boston is the clearest case — seed-1 alone would report SB as mildly *better*, while the 100-seed paired Wilcoxon test rejects with $p = 5 \times 10^{-4}$ in the *opposite* direction. We provide this comparison not to make a critical claim about prior reporting practice in the OOB literature, but to illustrate why second-order diagnostics in bagging studies require replication budgets well beyond three seeds.